\documentclass[utf8]{frontiersSCNS}

\usepackage{url,hyperref,lineno,microtype,subcaption}
\usepackage[onehalfspacing]{setspace}
\usepackage{mymacros}

\def\keyFont{\fontsize{8}{11}\helveticabold }
\def\firstAuthorLast{Bannikova \& Sergeev} %use et al only if is more than 1 author

\def\Authors{Elena Yu. Bannikova\,$^{1,2}$ and Alexey V. Sergeyev\,$^{1,2}$}

\def\Address{$^{1}$Institute of  Radio Astronomy, National Academy of Science, Ukraine \\
$^{2}$V.N. Karazin Kharkiv National University, Ukraine}

\def\corrAuthor{Elena Yu. Bannikova}
\def\corrEmail{bannikova@astron.kharkov.ua}

\begin{document} 
\onecolumn
\firstpage{1}

\title[Dynamics and formation of tori in AGNs]{Dynamics and formation of obscuring tori in AGNs}

\author[\firstAuthorLast ]{\Authors} %This field will be automatically populated
\address{} %This field will be automatically populated
\correspondance{} %This field will be automatically populated

\extraAuth{}% If there are more than 1 corresponding author, comment this line and uncomment the next one.
%\extraAuth{corresponding Author2 \\ Laboratory X2, Institute X2, Department X2, Organization X2, Street X2, City X2 , State XX2 (only USA, Canada and Australia), Zip Code2, X2 Country X2, email2@uni2.edu}

\maketitle

\begin{abstract}

 We considered the evolution of a self-gravitating clumpy torus in the gravitational field of the central mass  of an active galactic nucleus (AGN) in the framework of the N-body problem. The initial conditions take into account winds with different opening angles. Results of our N-body simulations show that the clouds moving on orbits with a spread in inclinations and eccentricities form a toroidal region. The velocity of the clouds at the inner boundary of the torus is lower than in a disk model that can explain the observed rotation curves. We discuss the scenario of torus formation related with the beginning of the AGN stage.

\tiny
\keyFont{ \section{Keywords:} Active Galactic Nuclei (AGN), quasars, Seyfert galaxies, N-body problems, Gravitation} 
\end{abstract}

\section{Introduction}
\label{sec:intro}
A dusty torus is an important structural element of an active galactic nucleus (AGN). In the framework of the unified scheme, the observational properties of AGNs (of type 1 and 2) are explained by the different orientation of the torus relative to an observer. This scheme was applied to Sy-galaxies \citep{Antonucci1993} and generalized to other classes of AGNs \citep{Urry1995}. Direct observations of obscuring tori exist only for a the nearby Sy-galaxies NGC 1068 \citep{Jaffe2004,Raban2009}, Circinus \citep{Tristram2007, Tristram2014}. 

The observed spectral energy distributions (SEDs) in the IR correspond to the model of a clumpy thick torus with a Gaussian distribution of clouds in its cross-section (Toroidal Obscuration Region) \citep{Nenkova2008a, Nenkova2008b}. The radiative transfer model of a clumpy torus with 3D cloud distributions was considered in \citep{Honig2010, Stalevski2012}. In the other radiative transfer models the IR emission is explained by two components: disk and wind (\citep{Honig2017} and references therein). ALMA observations allowed to estimate the mass of the torus in NGC 1068: $M_{torus}=10^5 M_{\odot}$ \citep{Garcia-Burillo2016}. It means that self-gravity of the torus can influence the motion of the clouds in it. In fact, the dynamics of the matter in the torus shows non-circular motions which can be related to its self-gravitating properties.

One of the main problem concerns the explanation of the geometrical thickness of 
the torus. In such a torus the vertical velocity component of the clouds must be 
of the same order of magnitude as the orbital velocity. Several mechanisms were 
offered for the solution of this problem. The geometrical thickness of the torus can 
be explained by IR radiation pressure \citep{Krolik2007, Chan2017, 
Dorodnitsyn2016}, by turbulent motions \citep{Schartmann2010}, or by starburst 
processes \citep{Wada2012, Wada2016}. Other models propose clumpy or dusty winds 
as an obscuring region (for example, \citep{Elitzur2006}). Indeed, many AGNs show 
the presence of outflows (winds) which can appear due to influence of the radiation pressure or of a magnetic field (\citep{Proga2004, Netzer2013}, and references therein).

Our main idea is that the geometrical thickness of torus can be achieved by the motion of clouds in inclined orbits \citep{Bannikova2012}. This assumption is quite natural because there is the external accretion of matter which replenishes the central region of AGN. On the other hand, since we consider a discrete medium in the torus, the orbital plane of each cloud must pass through the central mass. In this case, the toroidal structure may form due to the motion of the clouds in inclined orbits.
 
In our previous papers \citep{Bannikova2012}, we used special initial conditions (Keplerian torus) for the investigation of the torus evolution (see also \citep{Bannikova2015}). In a Keplerian torus, the distribution of particles by eccentricities and inclinations obeys some law and all major semi-axes equal a major radius of the torus. Here we continue to investigate the properties and stability of a self-gravitating clumpy torus in the framework of N-body problem for more general initial conditions, taking into account the wind cones with different opening angles. In Section \ref{sec:nbody} we discuss the initial conditions and present the result of N-body simulation.  Section \ref{sec:distribution} and Section \ref{sec:dynamics} are devoted to the main results on the cloud dynamics in the torus and to the conditions of obscuration (Section \ref{sec:obscuration}). A discussion of possible scenario of torus formation is provided in Section \ref{sec:discussion}.

\section{N-body simulation of a torus: initial conditions}
\label{sec:nbody}

\begin{figure}[h!]
\begin{center}
\includegraphics[width=16cm]{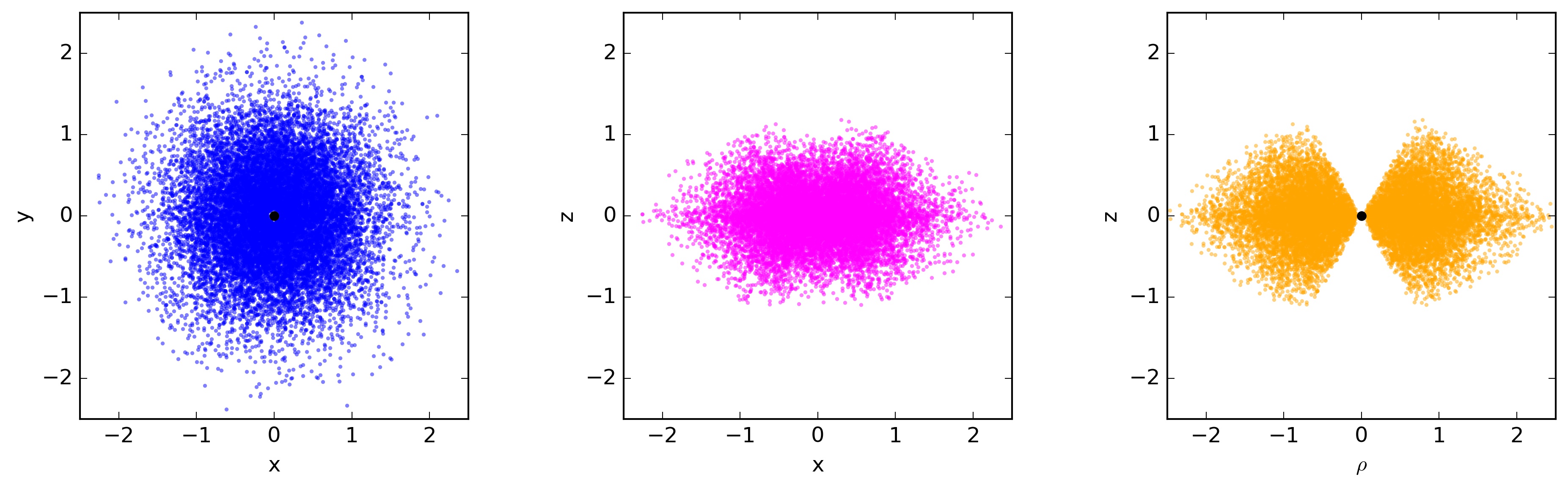} \\% This is a *.jpg file 
\includegraphics[width=16cm]{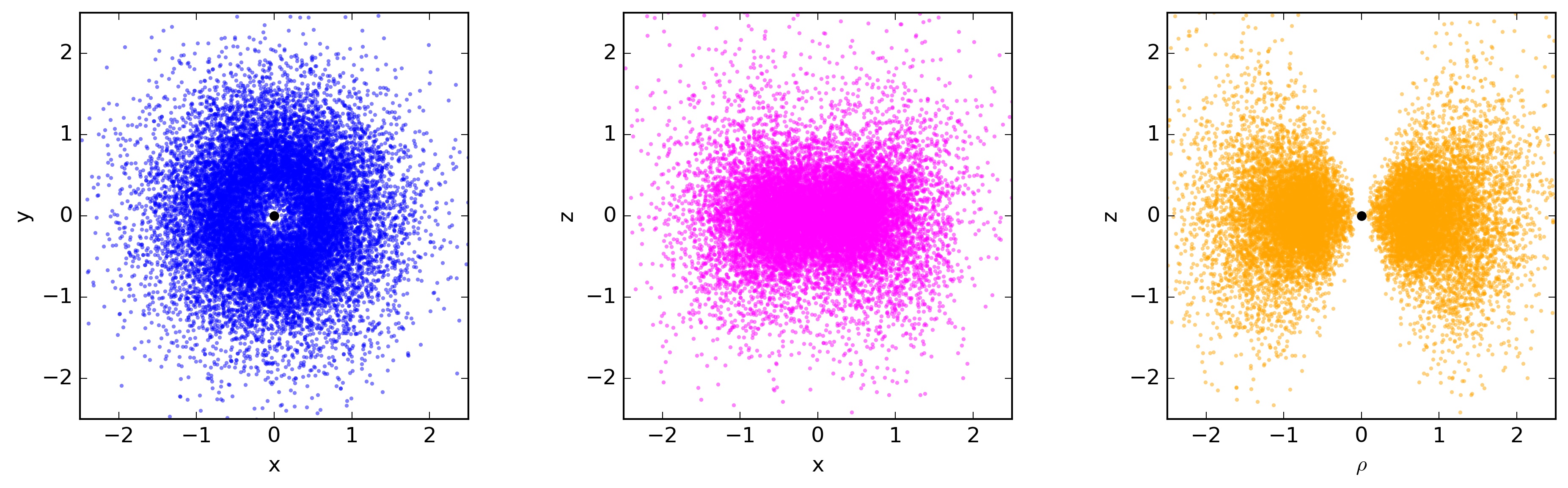}% This is a *.jpg file
\end{center}
\caption{
Distribution of the clouds in the torus for the initial state ({\it top 
panels}) and after $1,000$ average orbital periods ({\it bottom panels}): 
projections on the equatorial plane ({\it left}), on the meridional plane ({\it 
center}), and with all the clouds gathered in one meridional plane 
$\rho=\pm \sqrt{x^2+y^2}$ ({\it right}). Initial conditions: 
$M_{torus}=0.01 M_{BH}$, $N=16, 384$, $\epsilon=0.01$, $\theta_0=60^\circ$.
}
\label{fig:1}
\end{figure}

We consider, as initial conditions, the random distribution of clouds over all orbital elements: eccentricity, inclination, major semi-axis, and three angles. Such an initial distribution is more general than in (Bannikova et al. 2012). To form a toroidal structure, an anisotropy in the distribution of the clouds associated with winds in AGNs is needed. In these wind cones, the clouds acquire additional momentum due to radiation pressure and may overcome the gravitational forces of the central mass (and torus), leaving the system. This fact is accounted here by a simple assumption: the clouds from two opposite polar opening angles are excluded. In this case, the half-opening angle of the wind is a parameter influencing the resulting equilibrium cross-section of the torus and the obscuration condition. Three projections of the initial distribution of the clouds for the half-opening angle of the winds, $\theta_{wind}= 30^\circ$,  are shown in Fig.\ref{fig:1} ({\it top panels}). We will use in the following a value of the half-opening angle of the torus, $\theta_0= \pi/2 - \theta_{wind}$, and show the results of simulations for different values of $\theta_0$ in Section \ref{sec:obscuration}. 

N-body problem is reduced to the numerical integration of the equations of motion taking account the gravitational field of the central mass and ($N-1$) clouds of constant mass, $M_{cl}=M_{torus}/(N-1)$:
\begin{equation}
\textbf{a}_i = - \frac{G}{R^2}\left(M_{BH}\frac{\textbf{r}_i}{r^3_i}+ M_{cl} \sum_{j=1,j\neq i}^{N-1}  \frac{\textbf{r}_i-\textbf{r}_j}{\left(\left|\textbf{r}_i-\textbf{r}_j\right|^2+\epsilon^2\right)^{3/2}}\right),
\label{eq:1}
\end{equation}
where $\bf{r}_i=(x,y,z)$ is the vector of a cloud from the central mass normalized to the major radius of the torus ($R$); $\bf{a}_i$ is the vector of the acceleration of the $i$-th cloud acquired from all of the clouds of the torus $M_{torus}$ and from the central mass $M_{BH}$. A softening length $\epsilon$ in the N-body problem allows us to avoid unlimited increasing of the gravitational forces by collisions of particles and can be interpreted as the radius of the cloud $R_{cl}=\epsilon R$. We choose the physical parameters corresponding to the case of NGC1068: $M_{BH}=10^7M_{\odot}$ and $M_{torus}=10^5 M_{\odot} = 0.01M_{BH}$.
Since the method of parallel calculations is used, the  number of particle in N-body problem must be $N=2^n$. In presented simulations we adopt  $n=14$ (or $n=13$), thus the number of clouds comes $N_{cl}=N-1=16,383$ (or $N_{cl}=8,191$). 

N-body simulations were carried out using a technology of parallel calculations with GPU (CUDA). We used the Euler method with a step $0.001$ to solve equation (\ref{eq:1}).  In this case the total energy of the system $E$ is a constant with a good accuracy, $|E-E_0|=5\cdot 10^{-6}$, where $E_0$ is the initial value of the total energy. We use the unity system: $M_{BH}=G=R=1$.

The result of the simulation shows that the torus cross-section is changed and achieves its equilibrium state after a few hundred orbital periods. The distribution of clouds after $1,000$ orbital periods are shown in the bottom panels of Fig. \ref{fig:1}. (One orbital period corresponds $T=30,000$ years for $M_{BH}=10^7 M_{\odot}$ and $R=1\,pc$.) It is seen that the clouds are spread along $z$-axis as compared to the initial state, and that the cloud density increases towards the center of the torus cross-section. Indeed, the torus potential in the N-body problem might be divided into two summands  \citep{Bannikova2012}: a regular part which is related with a smooth potential of the torus, and an irregular one which is due to the gravitational interactions of the clouds. The regular potential leads to an increase of the cloud density towards the center of the torus cross-section, but the irregular forces between clouds tend to stretch it. As the result, the self-gravitating torus is geometrically thick, which is needed for the obscuration condition (see Section \ref{sec:obscuration}). 

We can suggest that the behaviour of the system for a larger number of clouds $N$ will not essentially differ from the obtained results. For example, if two toroidal systems have the same mass but different numbers of clouds ($N_1$, $N_2$), the same velocity dispersion of the clouds in the torus will be reached for the second system in the time interval $\bigtriangleup t_2= \bigtriangleup t_1 N_2/N_1$, where $\bigtriangleup t_1$ is the time interval for the first system \citep{Bannikova2012}.

\section{Distribution of the clouds in a torus}
\label{sec:distribution}

At equilibrium, the clouds are distributed in such a way that they form the toroidal structure: the number of clouds is exponentially decreasing along the $z$-axis (Fig.\ref{fig:1}, {\it bottom panels}), and the distribution of the clouds in the torus cross-section is Gaussian. This distribution is similar to that obtained by \citep{Nenkova2008a, Nenkova2008b} from the analysis of the spectral energy distribution (SED) in the IR and was also used in 3D radiative transfer model of a clumpy torus \citep{Honig2010}.   In our simulation, such a distribution is produced by the gravitational interaction between all the clouds and the central black hole.
\begin{figure}[h!]
\begin{center}
\includegraphics[width=16cm]{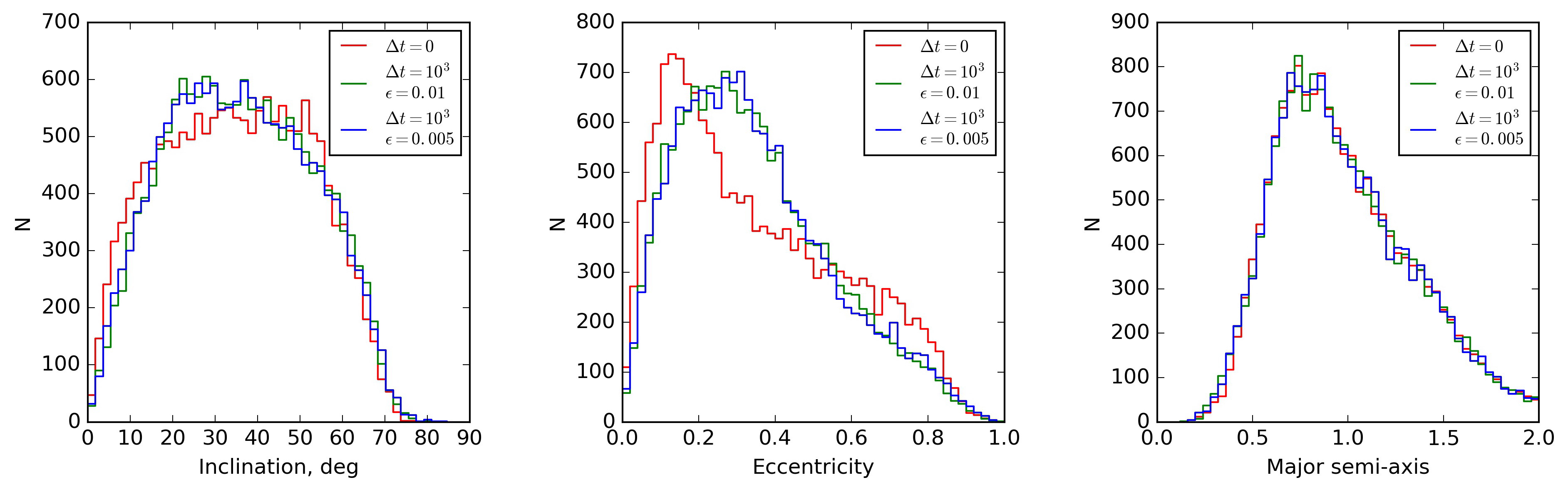}% This is a *.jpg file
\end{center}
\caption{Histograms of the orbital elements for the initial condition ({\it red}) and after $1,000$ average orbital periods for inclination ({\it left}),  eccentricity ({\it center}), and major semi-axis ({\it right}), for $\epsilon = 0.01; 0.005$ ({\it green, blue}). All other parameters correspond to Fig.\ref{fig:1}. }\label{fig:2}
\end{figure}

Knowing from N-body simulations the coordinates and velocity components, we can calculate the orbital elements of each cloud. Fig.\ref{fig:2} shows a comparison of the cloud distribution in the initial state ({\it red curves}) and after 1,000 average orbital periods for two values of the softening length $\epsilon=0.01$ ({\it green curves}) and $\epsilon=0.005$ ({\it blue curves}). It can be seen that the resulting distributions for the inclination and the major semi-axis (Fig. \ref{fig:2} {\it left}, {\it right}) do not essentially differ from the initial ones, while there is a little difference in the distribution for eccentricity (Fig. \ref{fig:2} {\it center}). This result shows that the initial state is near to the equilibrium one, and therefore, these distributions are statistically similar. Note that this torus evolution differs from the result presented in \citep{Bannikova2012}, where the initial condition (Keplerian torus) was far from equilibrium and the distribution of clouds was noticeably modified by the evolution of the system. It is seen  that the softening length does not influence the resulting distribution of the clouds because the equilibrium state, achieved due to self-gravity, depends on the mass (or the volume density) of the torus.      

\section{Dynamics of the clouds in dusty torus}
\label{sec:dynamics}

Rotation curves allow us to understand the dynamics of clouds in a dusty torus from observations of the megamaser emission. Indeed, the conditions for the formation of water maser emission ($\lambda = 1.35\,cm$)  appear at the inner region of the torus. Rotation curves were obtained only for a few nearby Sy-galaxies, including NGC1068 \citep{Gallimore1996, Gallimore2001, Gallimore2004, Greenhill1996}, and demonstrated that the matter in the torus is in sub-keplerian motion. Some models were proposed to explain such a motion and the rotation curve in NGC1068,  considering the torus in the approximation of a self-gravitating disk with the mass comparable to the central mass \citep{Hure2002, Lodato2003}. This does not agree with the mass of the torus obtained from ALMA observations, $M_{torus}=10^5 M_{\odot}$ \citep{Garcia-Burillo2016},  which corresponds to $0.01M_{BH}$. 
\begin{figure}[h!]
\begin{center}
\includegraphics[width=16cm]{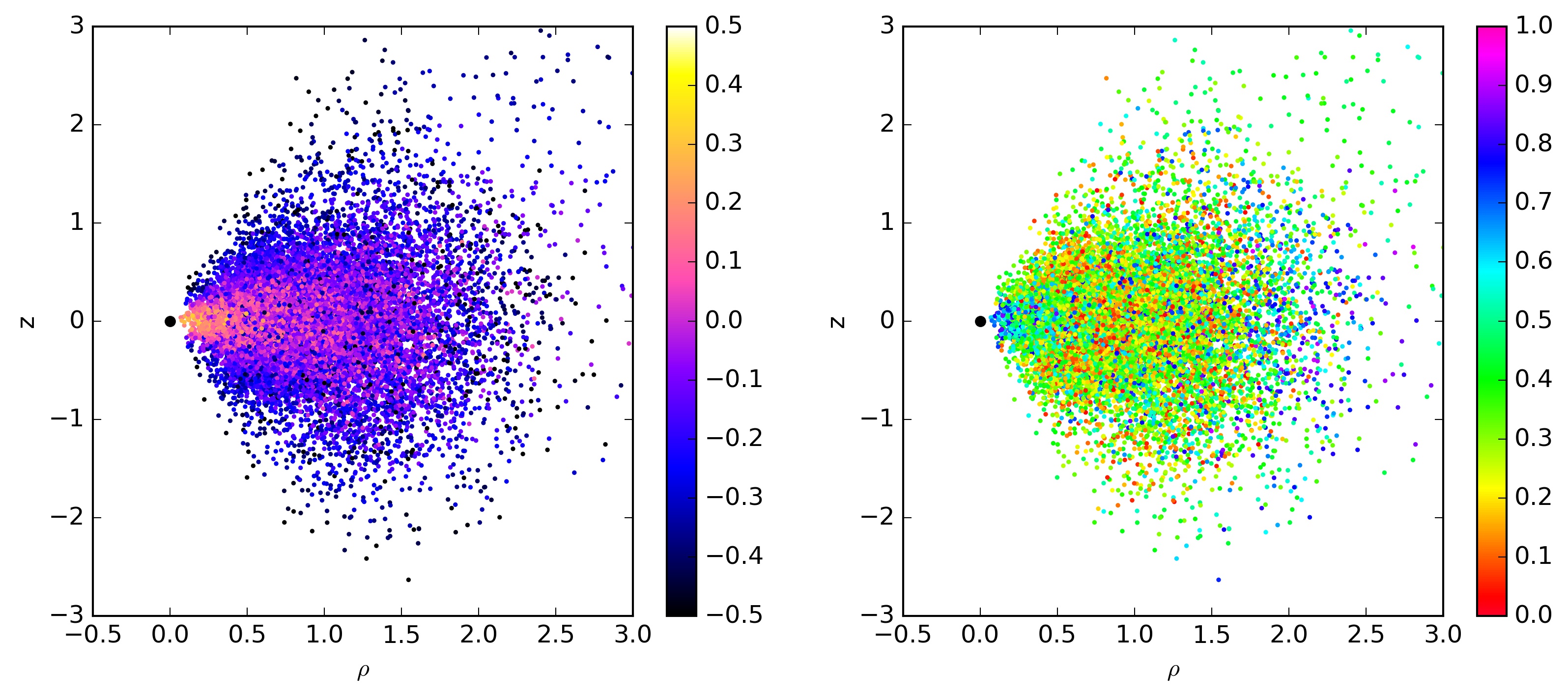}% This is a *.jpg file
\end{center}
\caption{Distribution of clouds in the torus. All particles are gathered in the meridional plane ($\rho$, $z$), where $\rho=\sqrt{x^2+y^2}$. {\it Left panel}: colors mark the deviations of the cloud velocity in the torus ($V$) from a keplerian velocity for the case of a disk ($V_k$): ($V/V_k-1$). {\it Right panel}:  colors mark the values of the eccentricity of the cloud orbits.}\label{fig:3}
\end{figure}

It is seen from Fig.\ref{fig:3} ({\it left}) that the motion at the inner boundary  of the torus ({\it black and blue points} at Fig.\ref{fig:3}, {\it left}) is sub-keplerian. Indeed, the inner boundary of the torus is formed by the clouds that move in orbits with different eccentricities and inclinations (Fig.\ref{fig:3}, {\it right)}. These clouds can pass through the apocenter or occupy some arbitrary (temporary) position on these orbits. A spread of the cloud velocities at the inner boundary of the torus can explain the observational data usually assigned to a turbulent motion. The clouds have high values of the velocity in the equatorial plane near the supermassive black hole ({\it orange points} at Fig.\ref{fig:3} {\it left}). These clouds move on orbits with large values of the eccentricity passing exactly through the pericenter; they could ultimately feed the accretion disk.

\section{Condition of obscuration}
\label{sec:obscuration}
Here we will determine the probability of obscuration of the central source by clouds of spherical shape with the radius $R_{cl}=\epsilon \cdot R$, where the major radius of the torus is $R=1$. It is convenient to use spherical coordinates, $(r, \theta, \phi)$, and consider a sphere of unit radius $r=R $. We divide the sphere by the coordinate $\theta$ into $n$ uniform parts (bins) whose length on the unit sphere of visibility is $h = \pi/n$.The ring length on the sphere for each angle $\theta_i$ is determined by the expression: $l(\theta) = 2\pi\cdot cos\,\theta$. Thus an area of a ring on the sphere can be written as:
\begin{equation}
S_{bin}(\theta)=l(\theta)\cdot h.
\end{equation}
Using the results of N-body simulations, we construct a histogram of the cloud distribution with respect to the angle $\theta$ between the equatorial plane and the line-of-sight with a cell size equal to $h$. The area of the projection of the cloud on the sphere is $S_i^{cl} =\pi (\epsilon/ {r_i})^2$. Let us suppose that the obscuration area equals the sum of all the cloud areas in the bin:
\begin{equation}
S^{cl}_{bin}(\theta)=\sum_{i}{S_i^{cl}}.
\end{equation}
Then we determine an obscuration coefficient as the ratio of these areas:
\begin{equation}
k_{obsc}=\frac{S^{cl}_{bin}(\theta)}{S_{bin}(\theta)}.
\end{equation}
\begin{figure}[t]
\begin{center}
\includegraphics[width=16cm]{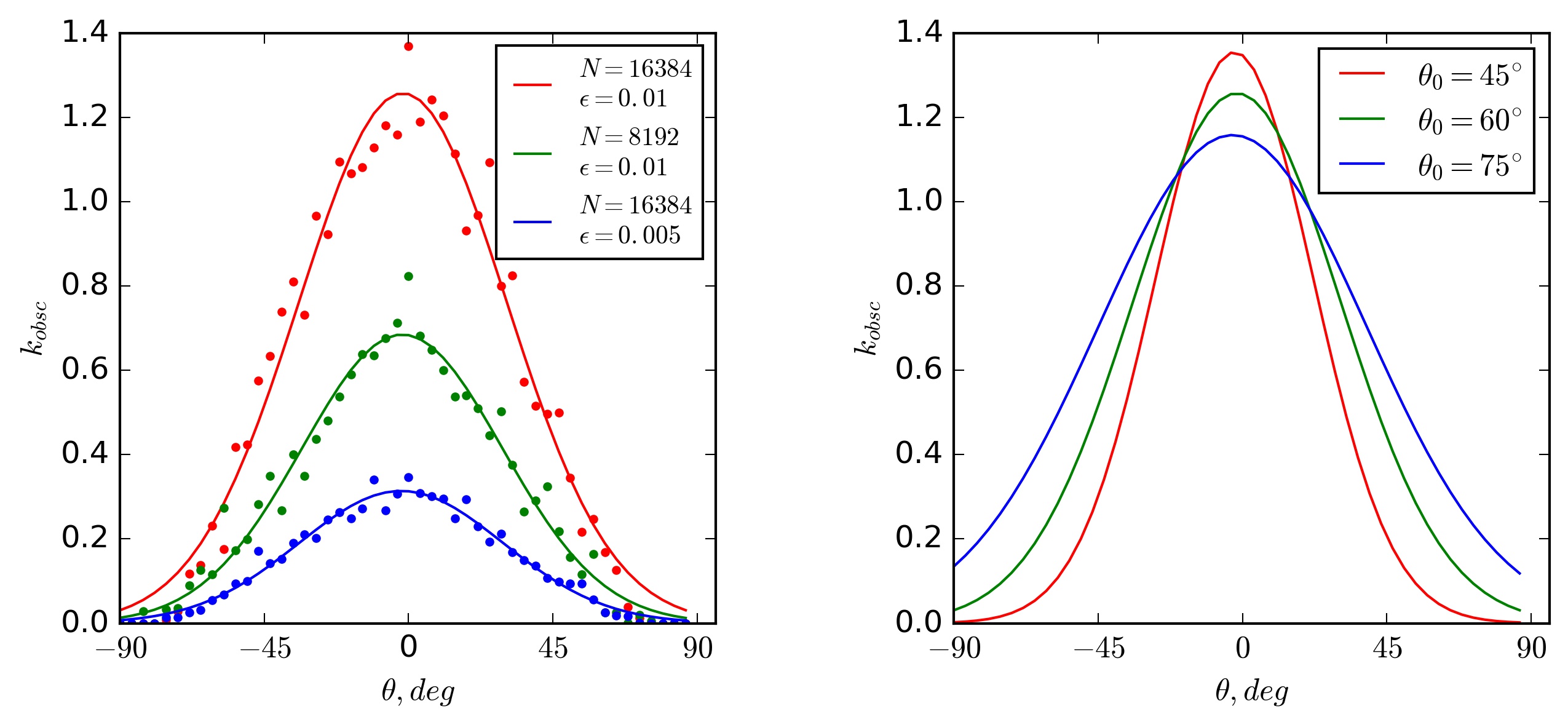}% This is a *.jpg file
\end{center}
\caption{Dependence of the obscuration coefficient on the angle between the line-of-sight and the equatorial plane. {\it Left panel}: values and the corresponding 
best-fitting curves are obtained from simulations for $\theta_0=45^\circ$, 
$N=16,384$ and $\epsilon=0.01, 0.005$;  $N=8,192$ and $\epsilon=0.01$. {\it Right 
panel}: each curve corresponds to the different initial half-opening angle of the 
torus: $\theta_0=45^\circ, 60^\circ, 75^\circ$.}\label{fig:4}
\end{figure}
It is seen from Fig. \ref{fig:4} that the obscuration coefficient depends on the inclination angle of the torus $\theta$ which is well fitted by the Gaussian function:
\begin{equation}
k_{obsc}(\theta)=A(N,\epsilon)\cdot
\exp\left(-\frac{\theta^2}{\sigma^{2}(\theta_0)}\right),
\end{equation}
where $\sigma(\theta_0)$ is the width of the torus half-opening angle, and the amplitude $A$ characterizes the number of the clouds in the equatorial line-of-sight for the certain value of $\epsilon$. Fig.4 shows that the amplitude $A$ has a linear dependence on the cloud number $N$ and a square dependence of the size $\epsilon$ (see also \citep{Bannikova2012}). The amplitude values for $N =16,384$ is $A = 1.26$, while for $N =8,192$ is $A = 0.68$. So, we can predict that for $N=10^5$ the number of clouds in the equatorial line-of-sight is about $7.7$ (for $\epsilon=0.01$), which is consistent with the estimations of radiative transfer models \citep{Nenkova2008a, Nenkova2008b, Honig2010}. 

The width of the Gaussian function, $\sigma$, depends on the initial half-opening angle of the torus. The right panel of Fig.4 shows the dependence of the obscuration coefficient on the angle between the line-of-sight and the equatorial plane. The fitting of function, $k_{obs}(\theta)$, gives the values of $\sigma(45^\circ)=24.5^\circ$, $\sigma(60^\circ)=32.4^\circ$, $\sigma(75^\circ)=41.9^\circ$, and the amplitude $A$ takes the values $1.35, 1.26$, and $1.16$.

\section{Discussion}
\label{sec:discussion}

N-body simulations show that a torus like the one observed in NGC1068 can stay thick, if its clouds initially have a random distribution of the orbital elements of the clouds and anisotropy in two polar directions. The clouds in such a toroidal structure move in inclined orbits with a spread in eccentricity, and the equilibrium state of the torus corresponds to Gaussian density distribution, which satisfies the obscuration conditions and the observed SED in the IR.  

The considered initial distribution of the clouds may be a consequence of the evolutionary processes in AGNs. It may suggest that, at the first stage, the supermassive black hole and the accretion disk are embedded in a quasi spherical distribution of dust \citep{Liu2011}. An example could be the system IRAS16399-0937, a galaxy whose core is immersed in quasi spherically distributed optically thick clouds \citep{Sales2015}. The beginning of the active stage may lead to an increase of the wind energy and to the anisotropy in the cloud distribution. Within the wind cones, the clouds acquire additional impulse against the gravitational forces due to radiation pressure. The dusty clouds located outside of the wind cones are unaffected by the wind and continue to move in inclined and eccentric orbits. These clouds can form the thick toroidal distribution (Fig. \ref{fig:1}, {\it bottom}) which plays the role of an obscuring structure in AGNs.

These simulations do not take into account the effects of dissipation, which will influence the distribution of the clouds and the stability of the torus. The dissipation can be related to the collisions of the clouds and their heating. More frequent collisions will occur near the center of torus cross-section, where density is higher. The heating of the clouds in this region can boost the radiation pressure which may be an additional factor compensating the dissipation effects. The clouds near the supermassive black hole will move to the center due to dissipation and eventually feed the accretion disk. The clouds moving in the inner (and outer) boundary of the torus will rarely collide because their number decreases exponentially. Thus, it can be assumed that the toroidal structure can be conserved for certain values of the dissipation coefficient. Note, that the main mechanisms which were proposed to explain the geometrical thickness of the torus may work in such a dynamical clumpy model. 

\section*{Acknowledgements}

We thank Massimo Capaccioli for the useful discussions, and the referees for helpful comments that improved the manuscript. 
One of us (EB) likes to thank the Organizing Committee of the conference "Quasars at all cosmic epochs" for financial support to present this work. 

\section*{Conflict of Interest Statement}

The authors declare that the research was conducted in the absence of any commercial or financial relationships that could be construed as a potential conflict of interest.
 
\bibliographystyle{frontiersinSCNS_ENG_HUMS}
\bibliography{bannikova} 
\end{document}